\documentclass[aps,prl,twocolumn,groupedaddress,showpacs]{revtex4}

\usepackage{graphicx}
\def\Dq{\delta_q}
\def\Tq{\theta_q}

\begin{document}

\preprint{UFIFT-AST-06-1}

\title{Can Electric Charges and Currents Survive in an Inhomogeneous
Universe?}

\author{E. R. Siegel}
\email[]{siegel@phys.ufl.edu}
\author{J. N. Fry}
\email[]{fry@phys.ufl.edu}
\affiliation{Department of Physics, University of Florida, %
Gainesville, FL 32611-8440}

\date{\today}

\begin{abstract}

Although observations point to the neutrality and lack of currents
on large scales in the universe, many mechanisms are known that
can generate charges or currents during the early universe.  We
examine the question of survivability of relic charges and
currents in a realistic model of the universe.  We show that the
dynamics of cosmological perturbations drive the universe to
become electrically neutral and current-free to a high degree of
accuracy on all scales, regardless of initial conditions.  We find
that charges are efficiently driven away in a time small compared
to the Hubble time for temperatures $100 \, \mathrm{GeV} \gtrsim T
\gtrsim 1 \, \mathrm{eV}$, while the same is true for currents at
all temperatures $T \gtrsim 1 \, \mathrm{eV}$.  The forced
neutrality relaxes constraints on the generation of an electric
charge in the early universe, while the forced erasure of currents
disfavors many mechanisms for the early origins of large-scale
magnetic fields.

\end{abstract}


\maketitle


The universe, on its largest scales, appears to be electrically
neutral and current-free.  Many scenarios exist, however, which
can lead to the production of either a net electric charge or an
electric current in the early universe.  The presence of either
charges or currents at early times has been discussed in the past
as possible candidates for generating seed magnetic fields in the
early universe, as well as causing other consequences for
cosmology.  In this {\it letter}, we expand upon our previous work
\cite{SF:06}, and demonstrate that although relic electric charges
and currents may reasonably exist at early times, they are driven
away in a realistic, inhomogeneous universe.

Many symmetries which appear to be good symmetries today may not
have been so in the past.  Such symmetries are often related to
conserved quantities, such as global symmetries (conserving baryon
and lepton numbers), or gauge symmetries (conserving color and
electric charges). Although no direct experimental evidence exists
for the violation of these conservation laws \cite{Eidelman:04},
the matter-antimatter asymmetry is a compelling indicator that
perhaps baryon number conservation was violated at some point in
the past \cite{Dine:03}.  Indeed, so long as the Sakharov
conditions \cite{Sakharov:67} are met, baryon number violation
appears to be quite likely.  Many mechanisms exist that could
explain the generation of a baryon asymmetry (baryogenesis),
including the decay of GUT-scale particles \cite{LKT}, sphaleron
processes \cite{tHooft:76} at the electroweak scale
\cite{Kuzmin:85}, sphaleron processing of a lepton asymmetry
\cite{FY:86}, or through a coherent scalar field \cite{AD:85}.

In a similar fashion to the global symmetry preserving baryon number,
it may be possible to break the gauge symmetry preserving
electric charge conservation.
These scenarios can result in a process
we refer to as {\it electrogenesis}, since they admit the
production of a net electric charge in the universe.
Perhaps the simplest pathway for electrogenesis is a theory where
the electromagnetic gauge symmetry $\mathrm{U}(1)_{\mathrm{em}}$
was temporarily broken in the past, only to be restored later at
lower temperatures. Langacker and Pi \cite{LP80} demonstrated this
possibility in the context of grand unification, while Orito and
Yoshimura \cite{OY85} and Nambu \cite{N84} showed that
electrogenesis is possible in higher dimensional theories such as
Kaluza-Klein. Another possibility that could lead to
electrogenesis is that the $\mathrm{U}(1)_{\mathrm{em}}$ symmetry
is not exact.  This arises in variable speed-of-light cosmologies
\cite{BLSV}, varying-$\alpha$ theories \cite{Shaw:06}, extensions
of the standard model with massive photons \cite{massphot},
brane-world models \cite{Dubovsky:00}, and models admitting
electron-positron oscillations \cite{M87}.

Additionally, electric currents can be generated at very early
times by many means, such as during inflation \cite{INF}, during
the QCD phase transition \cite{QCD,SOJ:97}, the electroweak phase
transition \cite{SOJ:97,BBM:96}, or by relativistic decays
\cite{DS93}. The earliest stages of structure formation also
generate small electric currents \cite{SF:06}, but these are too
small to be of any significance until much lower temperatures $(T
\ll 1 \, \mathrm{MeV})$.  Prior treatments of a universe with some
sort of electromagnetic asymmetry have assumed that charges and/or
currents within a comoving volume remain constant
\cite{LB59,OY85,DS93,SP96}. However, charge conservation still
allows the local charge to change, provided there is a flow of
current, and therefore there can be local changes in the
charge-per-baryon $(\Delta)$ at different epochs.  Observational
constraints on $\Delta$ at different epochs are as follows: the
anisotropies of cosmic rays place a constraint that at present, $|
\Delta | < 10^{-29} \, e$ \cite{OY85}; the isotropy of the
microwave background constrains $| \Delta | < 10^{-29} \, e$ at $z
\simeq 1089$ \cite{CF:05}; and big bang nucleosynthesis requires
that $ | \Delta | \lesssim 10^{-32} \, e$ at $z \simeq 4 \times
10^8$ \cite{MR02}.

In order for a local cosmological charge or current to have any
major significance, they must be correlated with the density
inhomogeneities which will grow into the collapsed structure
observed today.  It is expected that charge over- and
under-densities should possess the same types of inhomogeneities
as baryons \cite{DS93}, and their evolution should therefore be
calculable in the same fashion.  We have recently shown that
gravitational forces, in combination with Coulomb forces and
Thomson scattering, all impact the evolution of a charge asymmetry
\cite{SF:06}.  The remainder of this {\it letter} examines the
evolution of a relic electric charge or current, and studies the
associated cosmological ramifications.



A realistic cosmological model of the universe will necessarily
include inhomogeneities on both subhorizon and superhorizon
scales, as mandated by inflation.  An excellent treatment of the
linear evolution of these perturbations, including the photon,
neutrino, baryonic, and dark matter components, is given in Ma and
Bertschinger \cite{MaBert:95}.
While it is possible that a cosmological charge asymmetry could
manifest itself in exotic forms, these possibilities are
unsupported by experiment. This includes searches for fractionally
charged particles \cite{SP96,fracq}, or a difference between the
strength of the proton and electron charges \cite{MM:84}. We
therefore consider that if a cosmic charge asymmetry exists, it is
due to a difference between the number densities of protons and
electrons, following our previous treatment \cite{SF:06}.

We choose to work in the conformal Newtonian gauge, defined by the metric
\begin{equation}
\label{metric} ds^2 = a^2(\tau) [ -(1+2\psi) d\tau^2 + (1-2\phi)
dx^i dx_i]\mathrm{,}
\end{equation}
where $\phi$ and $\psi$ describe the scalar-mode inhomogeneities.
Including Coulomb interactions and Thomson/Compton scattering with
photons, the evolution equations for the proton and electron
fluids become
\begin{eqnarray}
\label{electrons}
\dot \delta_i &=& - \theta_i + 3 \dot \phi \mathrm{,}
\nonumber\\
\noalign{\smallskip}
\dot \theta_i &=& - \frac{\dot a}{a} \theta_i + c_s^2 k^2 \delta_i
+ k^2 \psi \\
&&{}+ \Gamma_i (\theta_\gamma - \theta_i)
+ \frac{4 \pi q_i e a^2}{m_e} (n_p - n_e) \nonumber \mathrm{,}
\end{eqnarray}
where $\delta_i$ is defined as the the departure of the local
spatial density of species $i$ from the overall spatial average
(i.e., $\delta_i \equiv \rho_i(x) / \bar\rho_i - 1$), and
$\theta_i$ is defined by $\theta_i \equiv i k^j (v_i)_j$, where
$v_i$ is the peculiar velocity of the field $\delta_i$. In
equation (\ref{electrons}), $i$ is $p$ for protons and $e$ for
electrons, $n_p$ and $n_e$ are the local number density of protons
and electrons, $q_p$ and $q_e$ are the proton $(+e)$ and electron
$(-e)$ charges, and $\Gamma_i$ is the (conformal time) rate of
momentum transfer due to photon scattering with charged particles,
\begin{eqnarray}
\label{Gammas}
\Gamma_e &\equiv& \frac{4 \bar{\rho}_\gamma n_e \sigma_T a}{3 \bar{\rho}_e}
\mathrm{,} \qquad \Gamma_p = \left( \frac{m_e}{m_p} \right)^3 \Gamma_e
\mathrm{.}
\end{eqnarray}
The Thomson cross section $( \sigma_T )$ for electrons is replaced
with the Klein-Nishina form for temperatures $T \gtrsim m_e$
\cite{B:74}.  By taking the difference between the density and
velocity fields for electrons and protons, a set of evolution
equations governing the evolution of a net charge and/or current
within a given volume is obtained. The equations for $\Dq$ and
$\Tq$, where $\Dq \equiv \delta_p - \delta_e$, $\Tq \equiv
\theta_p - \theta_e$, are
\begin{eqnarray}
\label{A}
\dot \Dq &=& - \Tq  \\
\dot \Tq &=& - \frac{\dot a}{a} \Tq + c_s^2 k^2 \Dq - \Gamma_e \,
(\theta_\gamma -\theta_b + \Tq) + \omega^2 \, \Dq  \mathrm{,} \nonumber
\end{eqnarray}
to linear order, where $ \delta_b =(m_p \delta_p + m_e
\delta_e)/(m_p+m_e) $ is the baryonic mass density perturbation
and $ \omega^2 = 4 \pi n_e e^2 a^2 / m_e $ is the (conformal time)
plasma frequency.  Note that other types of scattering, such as
Rutherford scattering, do not need to be included in equation
(\ref{A}) at this epoch, as the momentum transfer arising from
Thomson/Compton scattering is by far the dominant scattering term
in the early universe.

In our previous paper \cite{SF:06}, we investigated the effects of
the source term in equation (\ref{A}), proportional to $
\theta_\gamma - \theta_b $, in the absence of any initial charge
or current asymmetry. We found that a local charge asymmetry, of
order $ \Dq \sim 10^{-34} $ on $ \mathrm{Mpc}$ scales, is
generated near the time of recombination. In this paper, we study
the fate of an initial charge $(\delta_q)$ or current $(\theta_q)$
asymmetry. We note that since equations (\ref{A}) are linear,
these two cases evolve independently of one another. As the rates
$\Gamma_e$ and $\omega$ are typically much greater than the
expansion rate $\dot a / a$ before decoupling, and much greater
than the term $c_s^2 k^2$ on all but the smallest scales, the
latter terms can be neglected, and equations (\ref{A}) can be
simplified into the single equation
\begin{equation}
\ddot{\delta}_q + \Gamma_e \dot{\delta}_q + \omega^2 \delta_q = 0
\label{harmosc} \mathrm{,}
\end{equation}
where we remind the reader that $\dot{\delta}_q = - \theta_q$. An
initial asymmetry evolves as a damped harmonic oscillator, with
slowly varying coefficients. If the coefficients were constant, we
would have the usual solution, with two modes,
\begin{equation}
\label{solns} \delta_q = c_1 e^{s_1 \tau} + c_2 e^{s_2 \tau}
\mathrm{,}
\end{equation}
where $c_1$ and $c_2$ are undetermined constants, and
$s_{1,2}$ are the two roots of $ s^2 - \Gamma_e s + \omega^2 = 0 $,
\begin{equation}
s_{1,2} = - \Gamma_e/2 \pm \sqrt{\Gamma_e^2/4 - \omega^2}
\mathrm{.}
\label{s12}
\end{equation}
For $\Gamma_e / 2 > \omega$, the modes are a fast and a slow decay,
while for $\Gamma_e / 2 < \omega$, there is an oscillation
with a damped envelope.

The rates $\Gamma_e$ and $\omega$ in fact vary with expansion, as
$\Gamma_e \propto a^{-3}$ and $\omega \propto a^{1/2}$, and so
change slowly on the expansion timescale. As a result, the
arguments of the exponentials in equation (\ref{solns}) change
from $ s \, \tau $ to $ \int s \, d\tau $, as can be seen
straightforwardly by changing variables to the logarithm of the
asymmetry, $ X = - \ln( \Dq/\delta_0) $, where $ \delta_0 $ is the
initial asymmetry, and its derivative, the damping rate $ s =
\Tq/\Dq $. From equations (\ref{A}), these quantities satisfy
\begin{eqnarray}
\label{Xdot}
\dot X = s  , \qquad
\dot s =  - \frac{\dot a}{a} s + c_s^2 k^2
 + \omega^2 - \Gamma_e s + s^2 \mathrm{,}
\end{eqnarray}
where the rates $ \omega $ and $ \Gamma_e $ are again much greater
than the expansion rate $ H = \dot a/a $ and the term $ c_s^2 k^2
$ on cosmological scales.  We expect $ \dot s $ is also small,
since the rates $\Gamma_e$ and $\omega$ vary slowly. In this
approximation, the damping rate $s$ is a root of $ s^2 - \Gamma_e
s + \omega^2 = 0 $, as above, and
\begin{equation}
X = \int s \, d\tau  = \int da \, {s \over \dot a} =
\int {da \over a} \, {s \over H}
\mathrm{.}
\end{equation}
The quantity $ s/H $ represents the rate of logarithmic damping
per Hubble time, or damping per log expansion factor.

\begin{figure}
\includegraphics[width=\columnwidth]{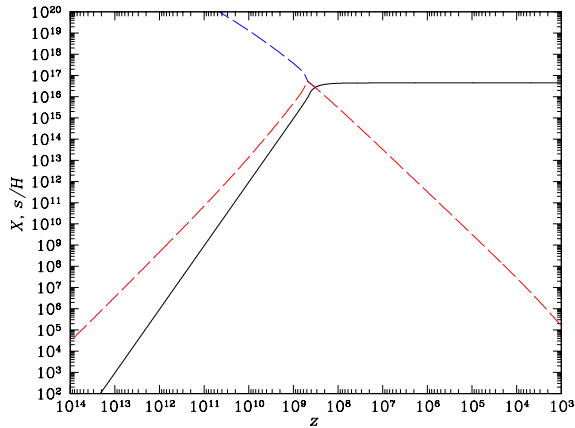}
\caption{The logarithm of the damping factor $ X = - \ln (
\Dq/\delta_0) $ (solid curve) and the normalized damping rate per
Hubble time $ s/H $ (dashed curves), as a function of redshift.
The lower dashed curve represents the slow decay mode, which
describes the evolution of an initial charge, while the upper
dashed curve illustrates the fast mode, describing the damping of
an initial current.  After critical damping $( z \approx 10^9 )$,
the two dashed curves evolve together. For damping rates $ s/H
\gtrsim 10^5 $, any initial charge or current will be completely
driven away in less than one Hubble time, which happens for excess
charges for values of $z$ between $10^3$ and $10^{13}$ and for
currents for $z \gtrsim 10^3$.} \label{fig1}
\end{figure}

Figure \ref{fig1} demonstrates dramatically that the damping is so
powerful as to wipe out a charge or current asymmetry for a large
window around the epoch of critical damping.
The figure shows the logarithm of the damping factor for both a
charge asymmetry $ \delta_q \propto e^{-X} $ and a current  $ J
\propto \theta \propto s \, e^{-X} $. At early times, an initial
charge asymmetry decays according to the slow-decay mode solution
of equations (\ref{Xdot}), while an initial current follows the
fast-decay mode. For values of $ X \gtrsim 10^2 $, an initial
charge asymmetry of order 1 is reduced to less than one excess
proton or electron per comoving horizon.

The results in Figure \ref{fig1} indicate that any excess charge
created at temperatures above $ 100 \, \mathrm{GeV}$ will be
entirely wiped out at roughly the electroweak scale, and any
electrogenesis that concludes between $T_{\mathrm{EW}}$ and
$T_{\mathrm{CMB}}$ will also be driven away in a time much less
than the Hubble time.  Furthermore, any relic current created at
temperatures $T \gg T_{\rm{CMB}}$ will be also driven away in a
time much less than the Hubble time. We therefore conclude that
the evolution of charged particles in the universe, including
Thomson/Compton scattering and the Coulomb interaction, force the
universe to an electrically neutral, current-free state,
independent of initial conditions, inflationary remnants,
electrogenesis, and phase transitions that may leave charges or
currents behind.

The results presented in this paper may have significant
implications for scenarios admitting electrogenesis in the early
universe.  The observational limits on the charge-per-baryon $(
\Delta )$ of the universe \cite{MR02,OY85,CF:05} have been used to
constrain models admitting electrogenesis in the past.  Our
results demonstrate that later-time constraints on $\Delta$ do not
constrain initial values of $\Delta$, as the universe drives away
any local charge asymmetries exponentially quickly.  Therefore, we
also conclude that the $ \mathrm{U}(1)_{\mathrm{em}} $ gauge
symmetry may have been broken in the past or may be only
approximately exact today, and the observed values of $\Delta$
would not be discernably different from zero.  Hence,
electrogenesis is not constrained at all by observational limits
on $\Delta$.  Nevertheless, the phenomenological consequences of
electrogenesis may leave signatures which can be searches for
experimentally.  Just as GUT-scale baryogenesis has
phenomenological consequences such as proton decay, electrogenesis
should also have consequences for particle physics.  Each specific
model that admits electrogenesis will have its own associated
phenomenology.  Some consequences of models admitting charge
nonconservation have been worked out \cite{CNC}, with various
distinct experimental signatures having been (unsuccessfully)
searched for \cite{exptCNC}.  The physical possibilities if either
Lorentz invariance or the $\mathrm{U}(1)_\mathrm{em}$ symmetry are
broken is quite rich and varied, and deserve further examination.

Also of cosmological importance are the implications for magnetic
fields in the early universe.  If large-scale currents are truly
driven away, as in Figure \ref{fig1}, then not only are magnetic
fields generated by a charge asymmetry \cite{DS93} ruled out, but
all magnetogenesis scenarios which rely on a relic current
persisting from early times are disfavored as well.  This requires
careful examination, as the scenarios for magnetic field
generation from the QCD \cite{QCD,SOJ:97}, electroweak
\cite{SOJ:97,BBM:96}, or inflationary \cite{INF} phase transitions
correctly discuss the growth of instabilities in the hot plasma of
the early universe. However, they do not examine whether the
currents which are initially generated by the phase transition
persist long enough to create such magnetic fields.  Our results
in Figure \ref{fig1} illustrate the damping rate for currents per
Hubble time, and we find the damping rate to be much quicker than
the growth rate of the instabilities given in the literature.  If
the currents do not persist long enough to create small-scale
fields, they will have no opportunity to then be transferred to
larger scales \cite{largeB}, creating seeds for the large-scale
magnetic fields observed today.  Since these currents do not have
a source driving them after their creation, the expansion dynamics
of the universe ought to damp them away very rapidly. Unless there
is some additional effect which prevents the currents that would
give rise to magnetic fields from being driven away, these
magnetogenesis scenarios are highly disfavored.

Since our results indicate that initial conditions on $\delta_q$
and $\theta_q$ are unimportant in the absence of sources, the most
significant contribution to a late-time $\delta_q$ or $\theta_q$
must come from the source contribution in equations (\ref{A}).
After any initial charge is wiped out, $\Dq$ and $\theta_q$ evolve
according to the dynamics in equations (\ref{A}), where the source
term, $\Gamma_e ( \theta_\gamma - \theta_b )$, determines $\Dq$.
Results from our previous work \cite{SF:06} indicate that the
typical charge-per-baryon on the scale of the present horizon is
$| \Delta | \sim 10^{-46} \, e$, a result that ought to be valid
since the present horizon is still in the linear regime of
structure formation.
Strictly, these calculations apply to asymmetries within the
horizon and at finite Fourier wavenumber $k$.  The asymmetry must
be within the horizon for its dynamics to evolve according to our
analysis due to causality; thus our results are definitely valid
for all wavenumbers $k$ from the moment they first enter the
horizon.  We also consider a na\"{i}ve (and not rigorous)
extrapolation to superhorizon $k$s. The dependence of our results
on $k$ is unimportant and vanishes as $ k \to 0 $. Thus, it
appears that even a global charge asymmetry may be driven to zero.
At first glance, this may seem to violate the principle of charge
conservation, but we recall that currents carry away excess charge
while still satisfying the continuity equation. An intuitive
picture is to consider an expanding universe containing
homogeneous proton and electron fluids free to expand at different
rates.  In the Newtonian limit, with spherical symmetry, an excess
of positive charge increases the expansion rate for protons and
decreases the expansion rate for electrons, and the total charge
in a given volume can decrease even though charge is locally
conserved. This construct does not work in detail, as different
expansion rates require a radial flow of protons relative to
electrons, and such a universe could be isotropic about only one
center. The flow may also require relativistic velocities.
However, the mechanism can apply on arbitrarily large finite
scales, and to reduce a charge asymmetry of order the mass density
inhomogeneity $( 10^{-5} )$ over the scale of the horizon requires
a relative velocity of $\mathcal{O}( \mathrm{km/s} )$, or a bulk
fluid displacement of $ \lesssim 1 \, \mathrm{Mpc}$.

Finally, although our calculations have accounted for high-energy
effects in electron-photon scattering \cite{B:74}, we have not
done this for protons.
We note that protons were dissociated into a quark-gluon plasma at
temperatures above the $\Lambda_{\mathrm{QCD}}$ scale, resulting
in different scattering rates and momentum transfer at these
energies.  Furthermore, we have not taken into account possible
shielding effects on cosmological scales, which, if present, could
cause the effective Coulomb force to fall off faster than $1 /
r^2$ for sufficiently large distances. Inclusion of these effects
may be needed to extract the exact behavioral details of
$\delta_q$ and $\theta_q$ at high energies.

The overall conclusion which can be drawn from this work is that
any relic charge or current created on subhorizon scales in the
early universe (at temperatures $T \gg 1 \, \mathrm{eV}$) is
driven away by cosmological dynamics.  This indicates that a net
charge or current of any magnitude could be generated at the QCD
phase transition, the electroweak scale, the time of Higgs
symmetry breaking, the time of supersymmetry breaking, the end of
inflation, or the grand unification scale, and the universe would
not be discernibly different from a universe that was electrically
neutral and current-free at all times.  This disfavors many
scenarios for the generation of seed magnetic fields, which rely
on the survival and sustainability of currents at early times. The
possibility that the universe underwent electrogenesis at some
point is intriguing and rich with possibility, and present
observations of electric charge neutrality do not preclude the
possibility that the universe (or a portion of it) was charged in
the past. Although the phenomenological consequences of
electrogenesis have been elusive thus far, evidence for its
existence would profoundly alter our cosmological picture of the
universe.

\begin{acknowledgments}
We thank Charles Thorn for useful discussions concerning gauge
symmetries and Jim Reardon for discussions on plasma physics, as
well as Karsten Jedamzik, G\"{u}nter Sigl, and Tetsuya Shiromizu
for their comments on the first draft of this paper. E.R.S.
acknowledges the Institute for Fundamental Theory at University of
Florida for support during the completion of this work.
\end{acknowledgments}


\end{document}